\documentstyle[12pt]{article}

\def\Pomeron {{\cal P}}
\def\SppS{Sp$\bar{{\rm p}}$S}

\def\thefigurelist#1{\section*{Figure Captions}
 \list {[\arabic{figure}]}{\settowidth\labelwidth{[#1]}\leftmargin\labelwidth
 \advance\leftmargin\labelsep
 \usecounter{figure}}
 \def\newblock{\hskip .11em plus .33em minus .07em}
 \sloppy\clubpenalty4000\widowpenalty4000
   \sfcode`\.=1000\relax}
 
\def\figitem#1{\refstepcounter{figure}\label{#1} \item[Fig.~\ref{#1}]}

\begin{document}
\begin{titlepage}

\begin{flushright}
   PSU/TH/116
\end{flushright}

\begin{center}
  {\large \bf Diffractive Hard Scattering with a Coherent
               Pomeron}
    \vskip 0.60in
        {\bf John C. Collins$^{a}$, Leonid Frankfurt$^{b}$,
             Mark Strikman$^{a}$}\\[0.1in]
       {$^{a}$Penn State University, 104 Davey Lab, University Park,
         PA 16802, U.S.A.}\\
        {$^{b}$Physics Department, University of
        Washington --- FM-15, Seattle, WA 98195, U.S.A.}\\
 \end{center}
 \vskip 0.5in

 \begin{abstract}

    Diffractive scattering involves exchange of a Pomeron to
    make a rapidity gap.  It is normally assumed that to get
    a hard scattering in diffraction, one may treat the
    Pomeron as an ordinary particle, which has distributions
    of gluons and quarks.  We show that this is not so:
    When we use perturbative QCD, there is a breakdown of
    the factorization theorem. The whole Pomeron can
    initiate the hard scattering, even though it is not
    point-like.  Qualitatively, but not quantitatively, this
    gives the same effect as a delta function term in the
    gluon density in a Pomeron.

\end{abstract}

\end{titlepage}


\section {Introduction}

Diffractive scattering in a high energy hadron collision
is a rather spectacular process.  One of the
initial hadrons emerges from the collision unscathed and
only slightly deflected, and there is a large rapidity gap
between the diffracted hadron and the other final-state
particles.  At collider energies,
10\% to 20\% of all interactions are
diffractive \cite{UA4,O}.
Diffraction is due
to Pomeron exchange, Fig.\ \ref{fig:diff}; but the exact
nature of the Pomeron in QCD is not yet elucidated.

Now, consider diffractive hard scattering; this means that we
consider diffractive final states that contain the products
of a hard scattering.
One example is the production of jets
in diffractive hadron-hadron collisions, for which data has
been reported \cite{Bon,Br} by the UA8 experiment.
Also, a few years ago, UA1 reported a
measurement \cite{Eggert} of diffractive bottom quark
production.

The conventional picture of hard scattering in
diffractive processes, due to Ingelman and Schlein
{}\cite{IS}, is to apply the usual hard-scattering formalism
{}\cite{fact} but to a Pomeron target.  For example, one has
deep-inelastic lepton-Pomeron scattering and jet production
in Pomeron-hadron collisions.  This leads to the concept of
the distributions of quarks and gluons in a Pomeron.  With a
typical ansatz for these distributions, one gets cross
sections of the correct order of magnitude \cite{Bon,Br}.

Frankfurt and Strikman \cite{FS} (see also Frankfurt
\cite{F})
showed that there are leading twist contributions
that do not correspond to the Ingelman-Schlein
mechanism.  These give what we will call coherent
diffractive
hard scattering, where almost all the longitudinal momentum
of the Pomeron goes into the hard scattering.  The latest
UA8 results \cite{Br} indicate that about 30\% of their
observed diffractive 2 jet events are consistent with
coherent hard diffraction.

In this paper we will elaborate the theory of coherent hard
diffraction by explaining
that the existence of leading twist coherent hard diffraction
is associated with a breakdown of the QCD factorization theorem
in diffractive processes.  Coherent hard diffraction is not
obtained from parton distributions in a Pomeron.

We will point out that one failure of the factorization
theorem is that we do not expect the same effect
to dominate in
diffractive
deep-inelastic leptoproduction.
Thus one would
not see universality of parton densities in the Pomeron when
one tries to make a simultaneous fit of the Ingelman-Schlein
formulae for diffractive deep inelastic lepton-hadron
scattering and for jet production etc in
diffractive scattering in hadron-hadron collisions.

In a recent interesting paper, Donnachie and Landshoff
\cite{DLHERA} have made a calculation
of diffractive production
of high $P_{T}$ jets in photon-proton collisions.
They find that because of a final state
interaction, they get events of a similar kind to our
coherent hard diffraction.  This is in a kinematic region
where the photon is point-like, so that there is not a
spectator jet for the photon.  Our prediction is that their
cross section is higher twist, in contrast to the leading
twist nature of coherent hard diffraction in proton-proton
collisions.
The explicit formulae given by Donnachie and
Landshoff are indeed higher twist, in agreement with our
general argument.  Thus their cross section
compared with the Ingelman-Schlein process decreases as
$1/P_{T}^{2}$ when $P_{T}$ is large.  Our coherent Pomeron mechanism
in hadron-hadron collisions is leading twist: it has no such
suppression at large $P_{T}$.


\section{Hard Scattering and Diffraction}


\subsection{Diffraction}

In diffraction, Fig.\ \ref{fig:diff}, we define $t$ to be
the invariant momentum transfer and $s'$ to be the mass
squared of
the final state excluding the diffracted proton.  Then we
let $\xi \equiv s'/s$, so
that $1-\xi $ is the fractional
longitudinal momentum of the outgoing diffracted
proton. We take $s$ to
be large, $-t \ll s$, and $\xi  \ll 1$. There will typically
be a rapidity gap of about $\ln (1/\xi )$ between the
diffracted proton and the rest of the final state.  In the
data from the UA8 experiment reported in \cite{Bon,Br,Sch},
$\sqrt s=630 \,{\rm GeV}$, and $1-\xi $ is between
about 0.90 and 0.96, so that $\sqrt {s'}$ is between
about 130 and 200 GeV.

The usual theory of diffraction uses Regge theory with a
triple Pomeron coupling to give a cross section
\begin{equation}
  s'\frac {d\sigma }{ds'\,dt}
  \simeq
   \left(\frac {s}{s'} \right)^{2(\alpha _{\Pomeron }(t)-1)}
   \left(\frac {s'}{s_{0}}\right)^{\alpha _{\Pomeron }(0)-1}
  \times  ({\rm function\ of\ }t).
\end{equation}
Here $\alpha _{\Pomeron }(t)$  is the Regge trajectory of the vacuum pole.
Collider data \cite{O,Sch,UA4} (from the ISR, the
{}\SppS{}and the Tevatron)
agrees with conventional expectations for this trajectory
(e.g., $\alpha _{\Pomeron }(0) \simeq 1.08$).

Because of the $t$ dependence of the Pomeron trajectory,
a softening of the distribution
over $\xi $ is predicted as $-t$ increases. This softening has
been recently observed at Tevatron \cite{O}. However, when
perturbation theory applies,
presumably at
large $-t$, QCD predicts $\alpha _{\Pomeron }(t)$ is larger than 1.
This is the first distinctive property of the perturbative
Pomeron.


\subsection {Hard Diffraction}

Now let us require that there be a hard scattering in
addition to the diffraction.  For example, we might ask that
there be a heavy quark, or a jet, or a Drell-Yan pair. Such
processes we will call `hard diffraction'.

We define $x_{A}$ and $x_{B}$ to
be the fractions of the initial hadrons' longitudinal
momenta that are carried into the hard scattering; these two
variables are defined at the parton level
for all processes with a hard
scattering.  Then we let $\beta $ be the fraction of the
Pomeron's momentum that goes into the hard scattering; this
satisfies $\beta =x_{B}/\xi $.
Since the variables $x_{A}$, $x_{B}$ and $\beta $
are defined at the partonic level, their values cannot be
readily extracted from an experimental event.  However, it
is possible to define from a 2 jet event, the minimum values
of $x_{A}$, $x_{B}$ and $\beta $ that are necessary to make the event
from parton-parton scattering.  When we discuss values for
these variables as computed from an event, it will be these
minimum values that we actually mean.

The model of Ingelman and Schlein is to use the usual
factorization formula but to apply it to Pomeron-hadron
scattering. This corresponds to saying that diagrams like
Fig.\ \ref{fig:IS} dominate, so that
\begin{eqnarray}
  \sigma (A + {\rm Pomeron} \to  {\rm Jets}+X)
   &=& \sum _{a,b} f_{a/A}(x_{A}) f_{b/\Pomeron }(\beta )
      \hat\sigma (a+b\to \hbox{Jets})
\nonumber\\
   && + \hbox{higher orders in $\alpha _{s}$}.
\end{eqnarray}
Here, $\hat\sigma $ is the short-distance parton-level cross
section for the hard scattering, which, for the sake of
definiteness, we have assumed to be for jet production;
it is the same as in totally inclusive hard scattering.
The sums are over all parton species (gluon, quarks and
antiquarks).
The densities, $f_{b/\Pomeron }(\beta )$, of partons in a Pomeron, should be
universal between different processes.  For example, the
same densities that are measured in diffractive
proton-antiproton collisions can be used to predict
diffractive deep inelastic scattering at HERA, if the
Ingelman-Schlein model is correct.


\subsection{Phenomenology}

One can estimate the size of the cross section
by using an ansatz for the distribution of partons
in a Pomeron.  The predictions are in at least
order-of-magnitude agreement with experiment
{}\cite{Bon,Br,Eggert}, when
the distributions are normalized by the momentum sum rule
(which is not necessarily valid \cite{DL} for the Pomeron).

Diffractive events will have the diffracted proton separated
from the rest of the event by a rapidity gap. This remainder
of the event, of mass $\sqrt {s'}$, is expected, in the Ingelman and
Schlein picture, to look qualitatively
like a normal hadron-hadron
scattering with $s$ set equal to $s'$.  (The detailed
properties of multiple Pomeron interactions can
change the quantitative properties of Pomeron-hadron
final-states as compared with hadron-hadron final states.)


\section {Coherent Hard Diffraction}

The Ingelman-Schlein model contributes to the cross section,
since the corresponding diagrams, Fig.\ \ref{fig:IS}, do
exist. We will now show that there is another contribution
where the whole Pomeron induces the hard scattering.  This
we term `coherent hard diffraction'.
The mechanism is illustrated by the lowest order graphs,
shown in Fig.\ \ref{fig:cohpombasic}.

There is an apparent similarity of Pomeron-hadron and
photon-hadron scattering, since Fig.\ \ref{fig:diff}
can also describe electron-hadron
scattering with exchange of a photon of low $q^{2}$. The
similarity is only apparent, since we will see that coherent
hard diffraction only occurs because of a breakdown of the
factorization theorem, whereas that theorem remains true for
photoproduction:  The photon can be genuinely
point-like, but the Pomeron
is not.


\subsection {Failure of Factorization Theorem for Coherent Hard
Diffraction}

\subsubsection {Leading graphs}

Let us consider the proof \cite{fact} of the factorization
theorem when one requires the final state to be diffractive.
To be definite, we will consider the hard
process to be the production of jets of some transverse
energy $E_{T}$.  The definition of the cross section will be
subject to some kinematic conditions, the exact details of
which will be irrelevant.  We will consider graphs at the
leading power of the relevant large or small variables
($E_{T}$, $s$, $\xi $), and we will focus particularly on the
region appropriate to coherent hard diffraction, i.e., where
$\beta \equiv x_{B}/\xi $ is close to unity.

The simplest graphs that give hard scattering
have one gluon coming into the hard
scattering from each hadron---Fig.\ \ref{fig:LOfact}.
These have a standard parton model interpretation:
\begin{equation}
   \sigma (A+B \to  {\rm jets}+X)  =
       f_{a/A}(x_{A}) f_{b/B}(x_{B}) \hat\sigma ,
\end{equation}
in terms of parton densities and a short distance cross
section.  The contribution of this graph is at the leading
power of the momentum scale, $E_{T}$, of the jet (for large
$E_{T}$).  There are power law corrections in $E_{T}$ (`higher
twist') to this result that are not covered by the
factorization theorem.

To get the leading power of $\xi $, for small
$\xi $, we need to exchange a line of the highest possible spin
between the lower proton and the hard scattering, that is,
we must have gluon exchange.
Since the effective spins of both the Pomeron and the
gluon are close to 1, the power law for $\xi $ is approximately
the same for gluon exchange as for diffraction caused by
Pomeron exchange.

However Fig.\ \ref{fig:LOfact} does not give a possible
model for coherent hard diffraction, since the gluon is a
color octet.  To get a diffracted proton with a rapidity
gap, we need color singlet exchange, which implies a minimum
of a two-gluon exchange, Fig.\ \ref{fig:cohpombasic}.

Individual graphs like Fig.\ \ref{fig:cohpombasic}
certainly give leading power contributions.  But they do not
have a parton model interpretation, since they do not
involve single gluon exchange.  To get the usual
factorization theorem, some kind of cancellation is
required.  These cancellations are part of the proof
\cite{fact}, in the case of a totally inclusive cross
section, with no diffractive requirement on the final
state.  If these cancellations were to remain true with
diffractive final states then coherent hard diffraction
would be a higher-twist effect.


\subsection {Soft Gluon Needed}

We will now show that the cancellations fail in the
diffractive case.  Now, both of the lower gluons in Fig.\
{}\ref{fig:cohpombasic} have transverse momenta much
less than $E_{T}$, the transverse energy of the jets.  Let
the longitudinal momentum fractions for the gluons be $x'$
and $\xi -x'$. There are two cases to consider: (i) Collinear
gluons: where both these momentum fractions are comparable.
(ii) Soft gluon: where $x' \ll \xi $ or $\xi -x' \ll \xi $.

Now, by gauge invariance, the single gluon in Fig.\
{}\ref{fig:LOfact} must be accompanied by a gluon field, and
this is accomplished by higher order graphs such as Fig.\
\ref{fig:cohpombasic}, with both the lower gluons in the
collinear region.  A relatively simple Ward identity
argument shows that multigluon exchanges in the collinear
region sum up to give a correctly gauge-invariant gluon
number density.  That is, we effectively get an exchange of
a single color octet gluon.  This part of the argument
depends on the fact that when the collinear gluons couple to
an oppositely moving gluon, as they do at the hard
scattering, then they form a highly virtual state.  The
virtuality is of order $E_{T}$.  This part of the argument is
unaltered by the requirement of a diffractive final state.

Soft gluons are another matter.  Graph-by-graph in Fig.\
\ref{fig:cohpombasic}, we have a leading contribution where
one of the exchanged gluons is soft, i.e., it has a very
small longitudinal momentum fraction.  In the graph, the
line marked `L' remains of low virtuality instead of being
off shell by order $E_{T}$.  This low virtuality line has a
rapidity that is on the opposite side of the rapidity of the
hard scattering to the diffracted proton.

Now, there is a very nontrivial cancellation of the effects
of
soft gluons in the inclusive cross section. The proof of the
cancellation is quite complicated {}\cite{fact}; it uses Ward
identities, causality, and a unitarity sum over final
states.

Part of the unitarity sum occurs where
the lower end of a gluon attaches to the diffracted proton
far in the future relative to the hard scattering.  Some
graphs involved in the cancellation are shown in Fig.\
\ref{fig:unitarity}.  These graphs cannot all satisfy the
diffractive condition.  Indeed, some of these final-state
interactions are relevant to filling in the
rapidity gap between the beam jets and the rest of the
event.  This part of the argument was known before the days
of QCD \cite{DEL,CW}, and it is used as part of the full QCD
proof \cite{fact}.

Some intuition can be gained from considering what happens
in space-time.  We have a collision of
Lorentz contracted and time-dilated hadrons.
At a particular point, jets are made by the collision over
a small distance scale $1/Q$ of
one gluon out of each hadron.
The
cross section is determined at the moment of the hard
scattering, and interactions outside the past light cone of
the hard scattering cannot affect the cross section.

But when we have hard diffractive scattering,
the remnants of one of the initial
hadrons must reassemble themselves into a single hadron
again, and that requires final-state interactions over a
long time scale.

Now the unitarity cancellation we need to get factorization
for the inclusive hard-scattering cross section involves
adding together collections of
graphs such as those shown in Fig.\
\ref{fig:unitarity}, which differ by the placement of the
final-state cut.  Some of these graphs involve octet
color exchange, and therefore do not contribute to
diffraction.  Non-factorization of the diffractive hard
scattering cross section is thereby established.  {\it The
non-factorization occurs at the lowest relevant order of
perturbation theory.}

Since the lowest order non-factorizing term comes from Fig.\
\ref{fig:cohpombasic}, essentially all of
the longitudinal momentum of the exchanged system goes into
the hard scattering.  Thus we have exactly the
effect advertised above.  There is an effectively coherent
contribution of the Pomeron to hard diffraction.

So far, we have considered the lowest order diagrams in pQCD for
hard diffraction.
These diagrams give the coupling for the perturbative pomeron to
the proton.  But, particularly at smaller $\xi $, we must
sum higher order graphs for the exchange.  Common wisdom is
that the sum of leading logarithms in $\ln \xi $ gives (in the
light cone gauge) a gluon ladder for the exchange of a
perturbative Pomeron. This ladder will not change the
coupling of the Pomeron, but it will change the power-law
dependence of the amplitude on $\xi $.  Such corrections become
very important when $\ln(0.1/\xi ) > 1/\alpha _{s}$.  Furthermore, extra
exchanged gluons can also contribute without power suppression.


\subsection {Power Laws}

A distinctive consequence of our factorization-violating
mechanism
is the dependence of the cross section on the variable
$\beta =x_{B}/\xi $. In the approximation
that radiation of extra gluons is
neglected, the diagrams of Fig.\ \ref{fig:cohpombasic}
correspond to a situation where the
whole momentum of pomeron initiates the hard scattering
\cite{F}.  Thus there is a $\delta (\beta -1)$ dependence on $\beta $.
Radiation of quarks and gluons will smooth out this delta
function.

A slightly different singular behavior, like $1/(1-\beta )$, is
obtained from diagrams
where two gluons (quarks)  are radiated from the same exchanged
gluon line \cite {FS}.  In both cases, we have a distinctive
sharp peak at $\beta =1$.
Such diagrams die out more rapidly with increasing $|t|$
than those of Fig.\ \ref{fig:cohpombasic}.

Moreover, a nontrivial change of the $\beta $ distribution
is expected as $t$ varies. According to conventional wisdom,
nonperturbative QCD physics dominates at small $t$. This is
the region where models of the Pomeron as a bound state of
quarks and gluons may be applicable.  Then the
Ingelman-Schlein mechanism would be appropriate.
We might expect the perturbative contribution to be
relatively small.  In particular, the coherent perturbative
$\delta (\beta -1)$ contribution is likely to survive only for small size
parton configurations in the diffracted hadron.

On the other hand, at larger $-t$, pQCD predicts
suppression of the long-distance nonperturbative physics,
as a result of the
color screening phenomenon.  The contribution of large
interquark distances in the wave function of
colliding protons  for diffractive processes is
suppressed, for the following reasons:
\begin{itemize}
\item[i]The lack of gluon radiation---for a
   recent discussion see \cite{S}.
\item[ii]Inelastic diffraction is suppressed for
   the collisions of black bodies.
\item[iii]The singularity of the potential in realistic quark
   models of a hadron leads to the dominance of small
   interquark transverse distances in the diffractively
   scattered proton when the momentum transfer $t$ is large.

\end{itemize}
Then the perturbative Pomeron should dominate, and coherent
hard diffraction should become much more important.  Thus
when $-t$ is around several GeV$^{2}$, a peak at $\beta =1$ should
become prominent.

We conclude that the $\beta $-dependence of hard diffractive processes
at  $\beta $ near 1 is a measure of the
perturbative Pomeron contribution at any $t$.
To visualize the expected dramatic effects, let us parameterize
the dependence of hard diffractive processes on $\beta $ by
the form $(1-\beta )^{N} $.
Typical models for soft physics give $N=5$
(multiperipheral models), or
$N=1$ (Pomeron as
bound system of gluons), etc.  But
the coherent pomeron contribution gives $N=-1$.
The relative size of the $N=-1$ term gives a measure of the
importance of perturbative Pomeron physics.

Another  distinctive feature of the perturbative Pomeron is
the dependence of the cross section for diffractive
dissociation on $\xi $.  Let us parameterize it as $1/\xi ^{n}$. When
$t$ is small, $n$ is less than 1 and decreases with $t$.
But when $t$ is large and $M_{X}^{2} \ll s$, we find that $n$ is
larger than 1, according to current wisdom of pQCD \cite{KFL}.

In reality, fully asymptotic behavior for the perturbative
Pomeron is only achieved at extremely large energies.
So the effective value on $n$ at practical values of $s$
will depend on $s$.  We expect that a reasonable description
of hard diffractive processes at the CERN collider and the
Tevatron, but possibly not in the SSC-LHC  energy range,
corresponds to  the exchange of two gluons and low-order
radiative corrections to this exchange.  (These radiative
corrections will give a few rungs in the ladder for the full
perturbative Pomeron.)

A further change in the
fraction of the coherent contribution
arises when $x_{A}$ is not small.
Now, the minimum value of $x_{A}$ is $4E_{T}^{2}/s'$, with $E_{T}$ being the
transverse energy
of one of the jets.
There is a factor in the cross section of $x_{A}G(x_{A},4E_{T}^{2})$ or
$x_{A}q(x_{A},4E_{T}^{2})$, depending on whether the Pomeron scatters off
a gluon or a valence quark.  Here $G(x_{A},Q^{2})$ and $q(x_{A},Q^{2})$
are the usual gluon and quark distributions in a
nucleon.  At small $x_{A}$, scattering off gluons dominates,
but at large $x_{A}$, scattering of quarks dominates.
The coupling of the perturbative Pomeron to a quark
is lower than its coupling to a gluon, and the
relative fraction of the coherent $\delta (x_{B}/\xi -1)$
term to the
Ingelman-Schlein continuum term will depend on the color
charge of
the parton coming from the non-diffracted hadron.


\section {Conclusions}

We have seen that there should be a coherent component of
hard diffraction in hadron-hadron collisions.  That is, the
whole Pomeron can induce the hard scattering.  The result is
that the products of the hard scattering (jets, or heavy
quarks, for example) will appear relatively close to the
edge of the diffractive rapidity gap.
Evidence for
a substantial coherent term in diffractive jet production
has recently been reported by the UA8 experiment \cite{Br}.
Further calculational work is needed to establish whether
the theoretical predictions are in quantitative agreement
with experiment.

The coherent contribution should be present in addition to
the contribution discussed by Ingelman and Schlein
\cite{IS}.  However, we should not expect factorization to
be exactly valid for the Ingelman-Schlein contribution.

At first sight, the coherent term behaves like a $\delta (1-\beta )$
component for the gluon distribution in a Pomeron.  (Compare
the experimental results from UA8 \cite{Br}.)  There will
also be
a singular $1/(1-\beta )$ contribution
\cite{FS}; this singular component will give similar
experimental consequences to the delta-function
contribution, after smearing by gluon radiation, by
hadronization and by calorimeter resolution.

But the rules of calculation are not those of standard hard
scattering, because of the need for a soft gluon.  This will
imply that we should expect violations of the universality
of the parton densities.

In particular, we should {\it not} expect the coherent
Pomeron to manifest itself in diffractive deep inelastic
lepton scattering, as at HERA.  The reason is that the soft
gluon in Fig.\ \ref{fig:cohpombasic} will no longer have an
opposite-side colored object to couple to.  We are talking
here about the kinematic region
where $\xi $, the fractional momentum of the
Pomeron, is not much larger than $x_{{\rm Bj}}$. But coherent hard
diffraction can be leading twist in the process $\gamma ^{*}(Q^{2})+
{\rm Pomeron} \to  {\rm jets}(E_{T}) +X$ when
\begin{equation}
  Q^{2} \ll E_{T}{}^{2} \ll s_{\gamma ^{*}+{\rm Pomeron}}.
\end{equation}
The higher twist result of Donnachie and Landshoff
\cite{DLHERA} applies when $E_{T}{}^{2}$ is of order $s_{\gamma ^{*}+{\rm
Pomeron}}$.

Another manifestation of the breakdown of factorization may
be that the ratio of coherent hard diffraction to
Ingelman-Schlein hard diffraction will vary in hadron-hadron
collisions.  Quark-initiated processes will have less of a
coherent fraction than gluon-initiated processes, because of
the smaller color charge of the quarks.  A comparison of
the Drell-Yan process with jet production, and of the cross
section at small $x_{A}$ and large $x_{A}$ would therefore be
rather illuminating.

Since we need separated color for our mechanism to provide
an effective contribution, we expect that Sudakov
effects can suppress it, just as in the Landshoff mechanism
for elastic hadron-hadron scattering. This is surely the
case for the
contribution of configurations with large  transverse distances
between quarks \cite{S}. But if we go to large
momentum transfer $-t$, then the transverse distances
involved should get smaller, where the Sudakov suppression
should be much reduced.  Note also that
the perturbative approximation to the Pomeron should be
much better than at low $t$.  The UA8 measurements have
moderate values of $|t|$---roughly between 1 and 2 GeV$^{2}$.

More work is obviously needed on quantitative estimates of
the cross section.

\section*{Acknowledgments}

This work was supported in part by the U.S. Department of Energy
under grant DE-FG02-90ER-40577, and by the Texas National
Laboratory Research Commission.  We would like to thank
colleagues for discussions, notably J.D. Bjorken, A. Brandt,
L. Lipatov, P. Schlein, D. Soper, and G. Sterman.



\begin{thefigurelist}{99}

\figitem{fig:diff} Diffractive scattering by Pomeron exchange.

\figitem{fig:IS}  Ingelman-Schlein model for hard
diffractive scattering.

\figitem{fig:cohpombasic} Typical lowest order graph for
coherent hard diffraction.

\figitem{fig:LOfact} Lowest order graphs for hard
scattering.

\figitem{fig:unitarity} Unitarity cancellation in hard scattering
   includes both of these graphs.

\end{thefigurelist}

\end{document}